# SPACE SOLAR APPLICATIONS
# Simulation with MATLAB- Simulink environment/Transmission of Solar Power Generated

Ferhat S. Demirci (1), Murari Koushik Muthu Kannapiran (2)

(1)(2) Illinois Institute of Technology,
Electrical and Computer Engineering, Graduate
South Federal Street, Chicago, IL 60616
USA

***Abstract*** **– In this paper a photovoltaic (PV) array suitable for use in outer space is simulated in the MATLAB/Simulink environment. The simulation of the PV array is first carried out with maximum power point tracking (MPPT) and then without MPPT. In a second step, the response of the system to different irradiation levels is observed. A brief overview of the transmission of solar power generated in space down to the earth via microwaves is also given. Selected simulation results are provided for a comparative study of both cases.**

## I. INTRODUCTION

Photovoltaic (PV) arrays have been an attractive research and application area for the last decade. PV units have been widely used to power small household devices and, in particular, industrial applications. Solar energy is one of the most effective, inexpensive, harmless and least polluting renewable energy sources. Its utilization can be categorized in two ways: solar heating/cooling and solar electricity. Solar energy can be converted into electrical energy by means of a PV array. PV systems are now well recognized and widely used in electric power technologies, and many applications have been developed, such as solar power generation, solar vehicles, battery charging, water pumping and satellite power systems.

Besides clean and safe generation of electricity, the required source is quite predictable. However, sunlight does not cover the whole 24 hours of a day, and atmospheric conditions decrease the effectiveness of the installed system. Additional algorithms, power electronic devices and mechanical structures are required to optimize the power extracted from sunlight. The aim of this paper is to illustrate the power extracted from a PV array located outside the earth's atmosphere. A further challenge is the transmission of the generated electricity down to the earth; this is reviewed briefly in Section III.

A solar power satellite (SPS) that transmits the solar power generated in space to the ground via microwaves is being investigated as a new energy technology. The PV model follows the approach in [16]; this paper adds the MPPT, current-controller and PWM implementation in Simulink and the simulations at space irradiance levels.

## II. APPLICATION TO PV SYSTEM DESIGN

First, the PV module for the space application is selected; its working conditions and efficiency are the two main selection parameters. The Emcore photovoltaic satellite solar panel is used in this paper.

MATLAB and Simulink form a core environment for Model-Based Design, in which mathematical models of the behavior of physical systems are built from predefined block-diagram elements that are dragged, dropped and connected. The resulting dynamic system models can be continuous-time, multi-rate discrete-time, or a combination of the two.

The cell selected for this application is the Emcore Triple-Junction Solar Cell for Terrestrial Applications (CTJ PV cell). It is a triple-junction cell on a germanium substrate with a multi-layer antireflective coating (low reflectance over 0.3–1.8 μm), weldable or solderable front and back contacts, and a standard aperture area of 1 × 1 cm². The cell is designed for terrestrial concentrated photovoltaic applications under high illumination. Its datasheet parameters are used here as an assumption for the simulated array; a space-qualified cell would have different parameters. The manufacturer reports a 4-inch Ge production capacity of more than 300 MW per year [1]. The specifications are given in Tables 1 and 2 below.

| | |
|---|---|
| External Dimensions | 10.68 x 10.075 mm |
| Designed Aperture Area | 10 x 10 mm |
| Total Nominal Active Area | 100 mm2 |
| Bondable Perimeter Busbar | 0.235 mm |
| Total Cell Thickness | ~0.2 mm |

Table 1: Standard CTJ Cell Dimensions

| | |
|---|---|
| Efficiency | 36.3% |
| Voc | 3.251 V |
| Isc | 16.37 A/cm2 |
| Vmp | 2.68 V |
| Imp | 16.04 A/cm2 |
| Pmp | 42.9 W/cm2 |

Table 2: Electrical Output Parameters

The operating point of the PV array is found by iterating the value of the current-controlled source according to the modeling error. The required PV voltage is calculated from the current-voltage equation:

$$V = (KT/q)\ln((Isc - I)/Io + 1) \qquad \text{Eqn.1}$$

$$e(k) = V_{required}(k) - V_{measured}(k) \qquad \text{Eqn.2}$$

$$I(k) = I(k-1) + \alpha e(k-1) \qquad \text{Eqn.3}$$

The error is calculated by subtracting the actual voltage from the required voltage. This error is then used to calculate the new current reference of the controlled current source. Where, $\alpha$ is the iteration coefficient. From the iteration equations it can be seen that as error goes to zero the current change decreases. This method is faster in terms of response and it is stable for all regions of the system. The series resistance is usually very small, for simplified simulation, this resistance can also be neglected. But for unstable operating case, the series resistance is important to damp the oscillations in current and voltage. Larger resistance value means better damping which also means high voltage drops for high current ratings. A very small resistance value may not damp the oscillations sufficiently.[16]

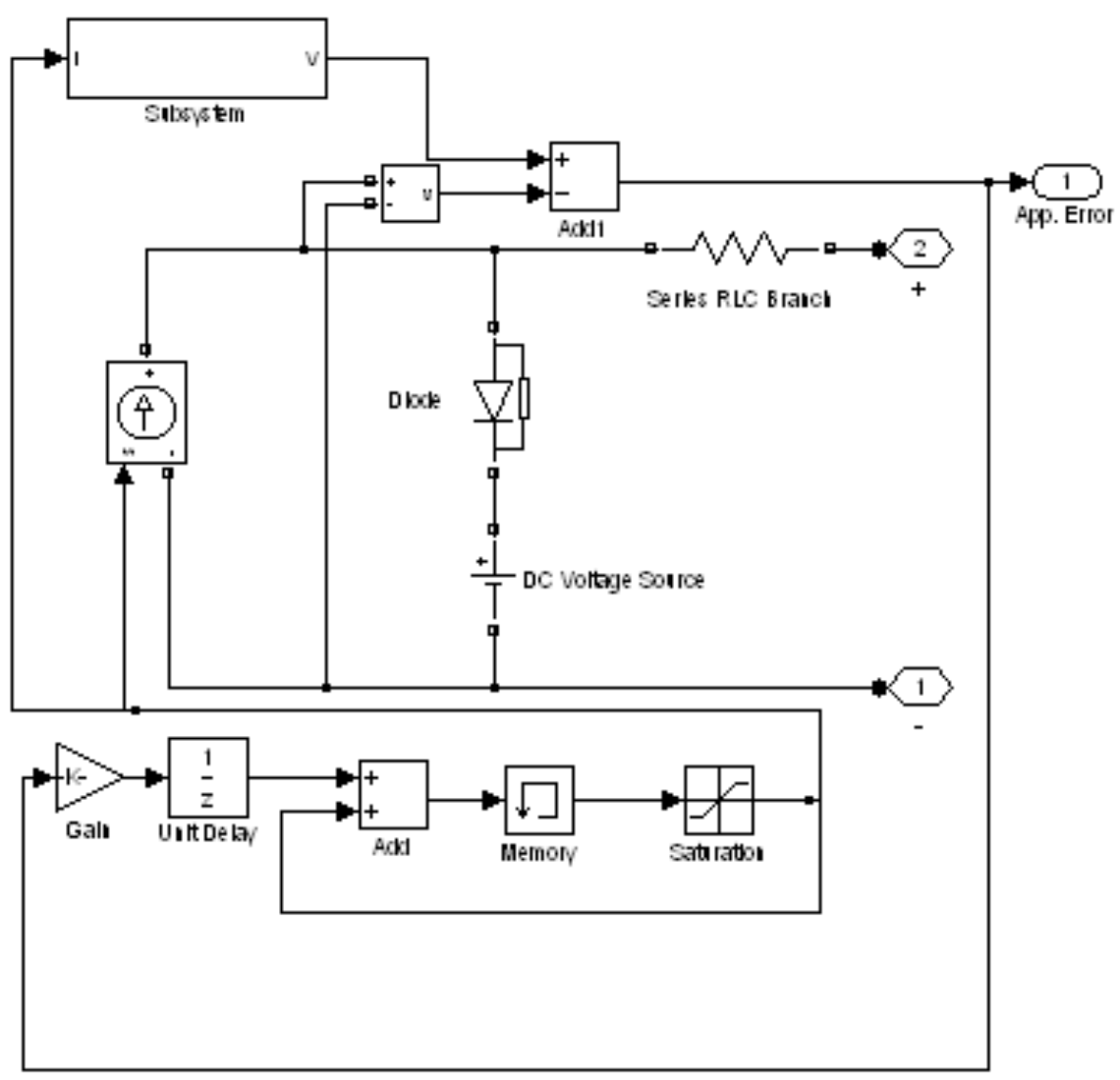

Fig. 1 PV Panel Design

Fig. 1 shows the PV panel design: a single-diode model of the PV array in which the parallel resistance is omitted. The parallel resistance of the single-diode model is normally large enough to act as an open circuit, so neglecting it has little effect on accuracy [16]. The net output current is the photocurrent Iph minus the diode current Id.

$$I_{PV} = I_{ph} - I_D = I_{ph} - I_0\left[\exp\frac{q(V_{PV} + I_{PV}R_s)}{\alpha kT} - 1\right]$$ Eqn.4

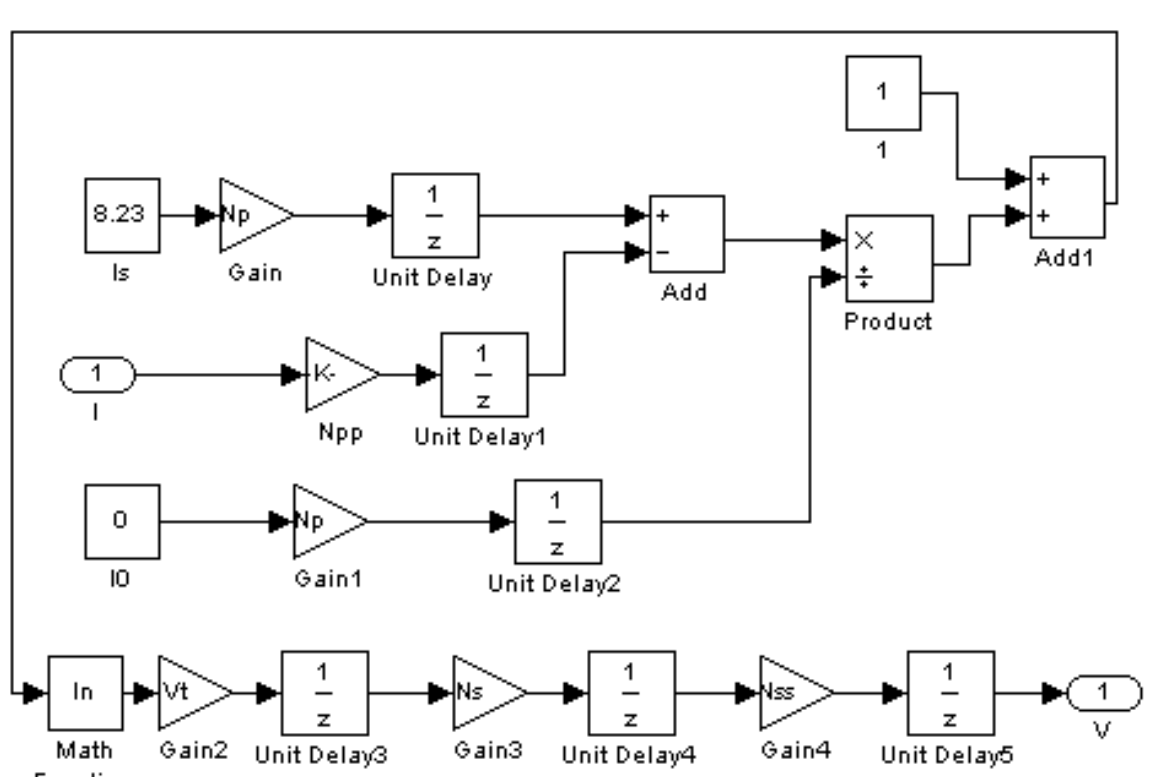

Fig. 2 Subsystem for Eqn.1

For the maximum power point tracking Incremental Conductance method is used in this paper. The incremental conductance [2, 3-13] method is based on the fact that the slope of the PV array power curve (Fig. 3) is zero at the MPP, positive on the left of the MPP, and negative on the right, as given by

$$\begin{cases} dP/dV = 0, \text{ at MPP} \\ dP/dV > 0, \text{ left of MPP} \\ dP/dV < 0, \text{ right of MPP} \end{cases}$$ Eqn.5

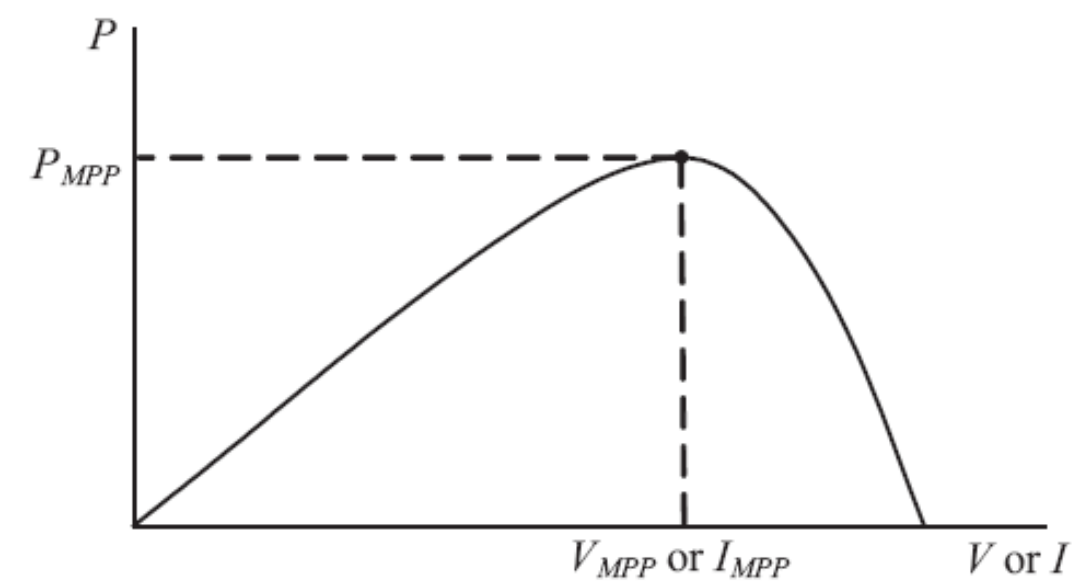

Fig. 3: PV Array Power Curve

Since

$$dP/dV = d(IV)/dV = I + V\,dI/dV \cong I + V\,\Delta I/\Delta V$$ Eqn.6

Eqn.5 can be written as follows;

$$\begin{cases} \Delta I/\Delta V = -I/V, \text{ at MPP} \\ \Delta I/\Delta V > -I/V, \text{ left of MPP} \\ \Delta I/\Delta V < -I/V, \text{ right of MPP} \end{cases}$$ Eqn.7

The MPP can thus be tracked by comparing the instantaneous conductance ($I/V$) to the incremental conductance ($\Delta I/\Delta V$) as shown in the flowchart in Fig. 4. $V_{ref}$ is the reference voltage at which the PV array is forced to operate. At the MPP, $V_{ref}$ equals to $V_{MPP}$. Once the MPP is reached, the operation of the PV array is maintained at this point unless a change in $\Delta I$ is noted, indicating a change in atmospheric conditions and the MPP. The algorithm decrements or increments $V_{ref}$ to track the new MPP.

The increment size determines how fast the MPP is tracked. Fast tracking can be achieved with bigger increments but the system might not operate exactly at the MPP and oscillate about it instead; so there is a tradeoff. References [8] and [12] propose a method that brings the operating point of the PV array close to the MPP in a first stage and then uses incremental conductance to exactly track the MPP in a second stage. By proper control of the power converter, the initial operating point is set to match a load resistance proportional to the ratio of the open-circuit voltage ($V_{OC}$) to the short-circuit current ($I_{SC}$) of the PV array. This two-stage alternative also ensures that the real MPP is tracked in case of multiple local maxima. Reference [14] uses a linear function to divide the I-V plane into two areas, one containing all the possible MPPs under changing atmospheric conditions. The operating point is brought into this area and then incremental conductance is used to reach the MPP.

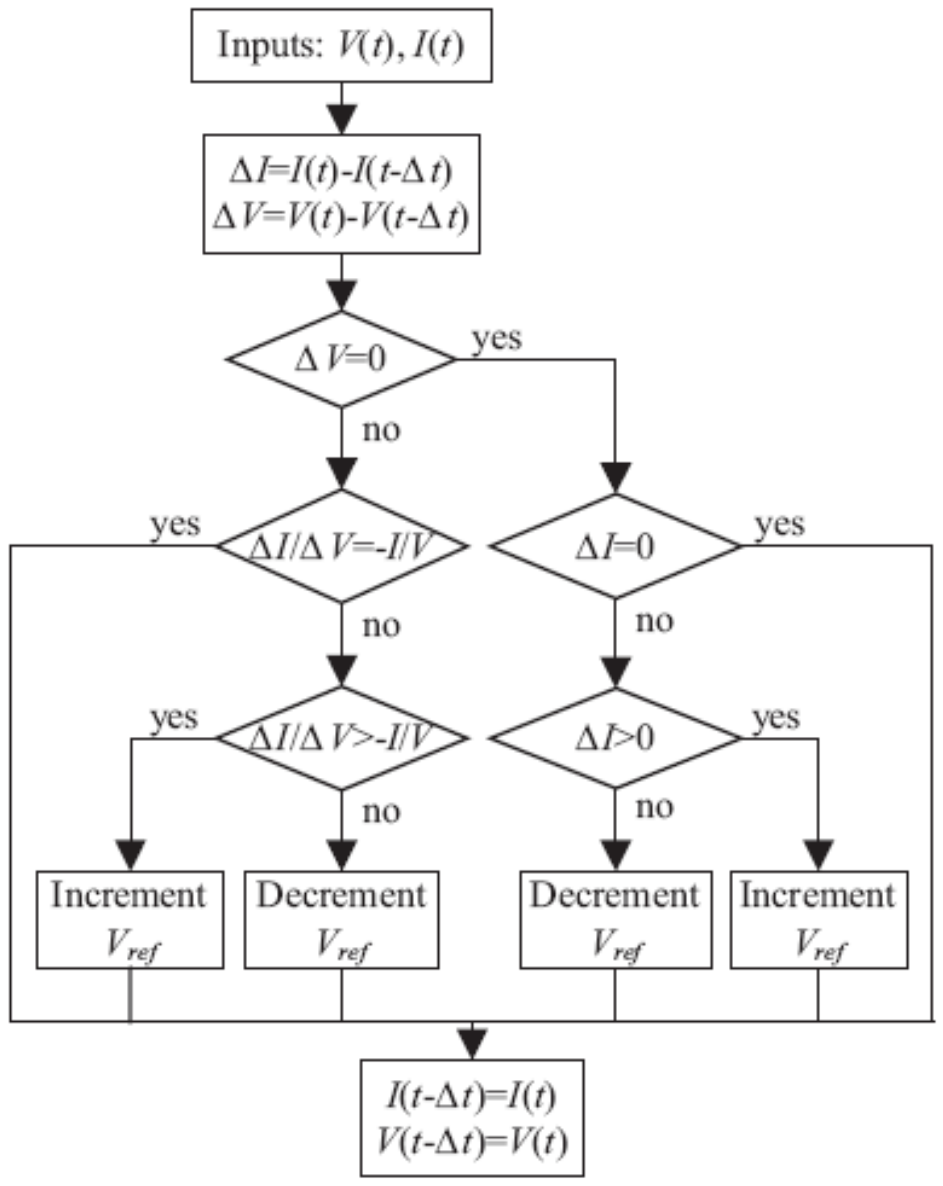

Fig. 4: Incremental Conductance Algorithm

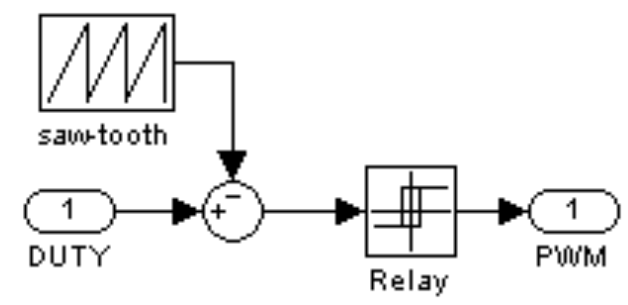

Fig. 7 PWM Generator

MPPT is composed of various stages such as;

- Maximum power point determination part which includes four subsystems in it for the determination of $\Delta I$, $\Delta V$, $-V/I$ and $\Delta V/\Delta I$. These subblocks can be seen in Fig. 8 and Fig. 9. For the maximum power determination part, Ipv and Vpv values are used. For the conditions of Incremental Conductance Method of $\Delta I=0$ and $\Delta V=0$ Fig. 8 is used. For $\Delta I\neq 0$ and $\Delta V/\Delta I=-V/I$ Fig. 9 is used.
- Current increment determination part can be seen in Fig. 5, which is used to control $\Delta I=0$, $\Delta V\neq 0$, $\Delta V>0$ and $\Delta I\neq 0$, $\Delta V/\Delta I\neq -V/I$, $\Delta V/\Delta I>-V/I$
- Calculation of current change, in this method the current change is based on the $\Delta V/\Delta I$ and $-V/I$. The advantage of using a variable change is that the controlling system responds faster and more accurately than a system using a fixed step change.

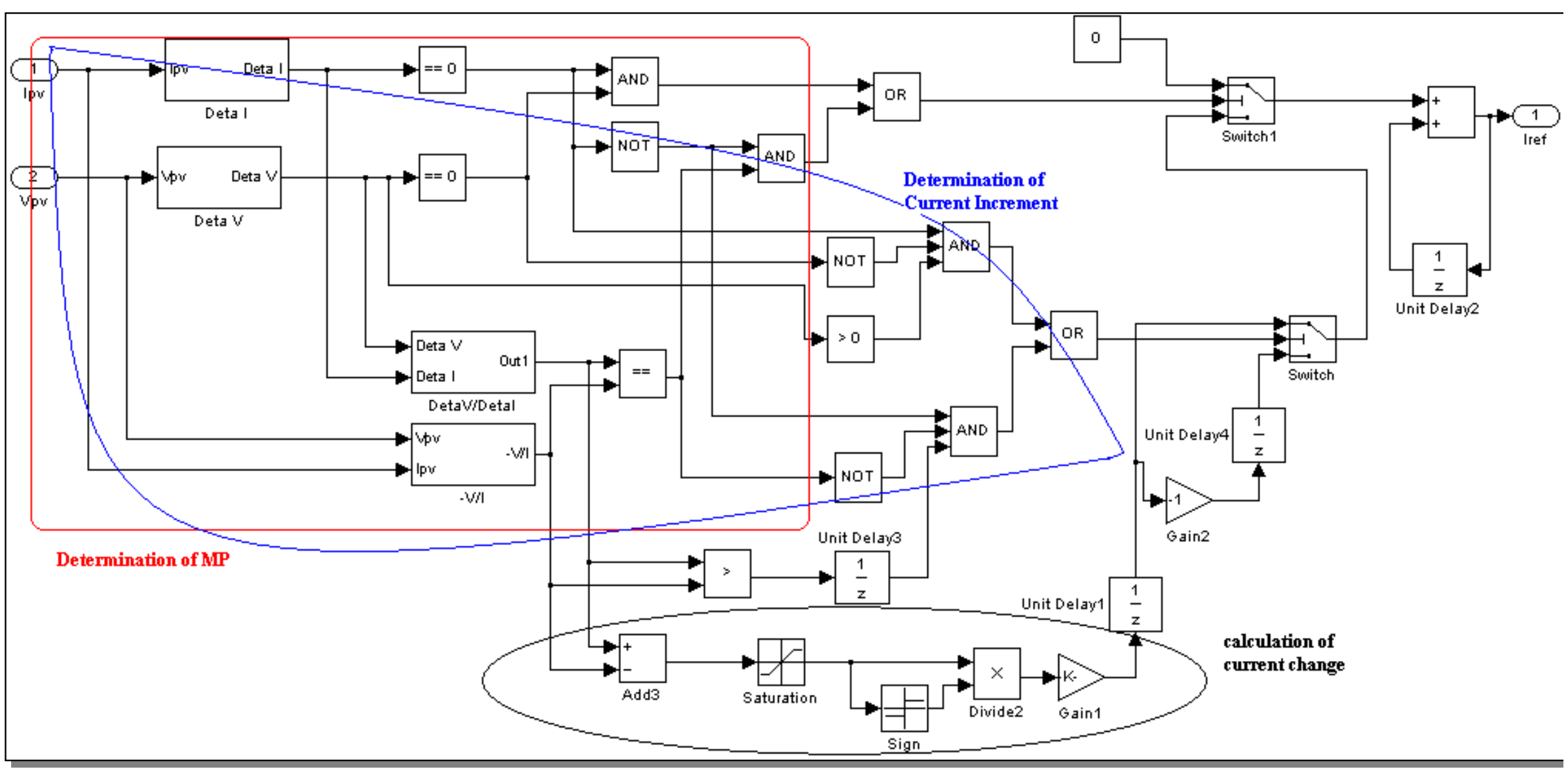

Fig. 5: MPPT Block

For the current controller in this project, difference between the real current and reference current has been taken into account and via PI control of this difference a duty cycle is obtained. Then this duty cycle is converted to the PWM signal. These application blocks are shown in Fig. 6 and Fig. 7. Switching frequency for the PWM generator is 10 kilohertz.

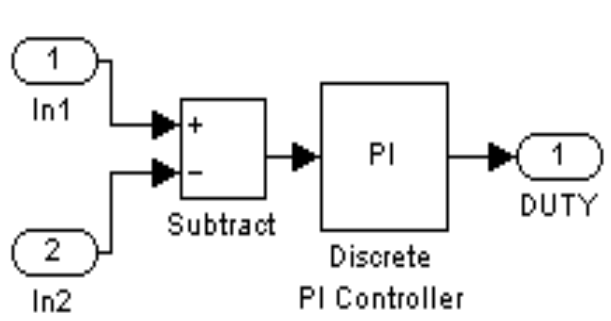

Fig. 6 Current Controller Block

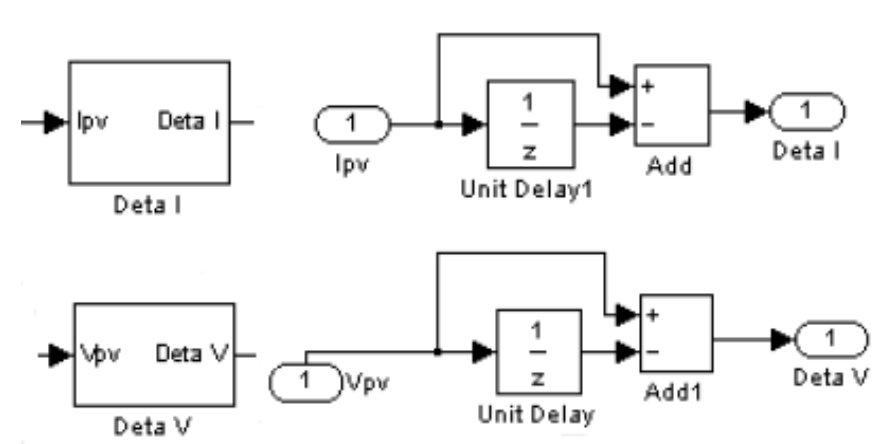

Fig. 8: Subblocks of ΔV and ΔI

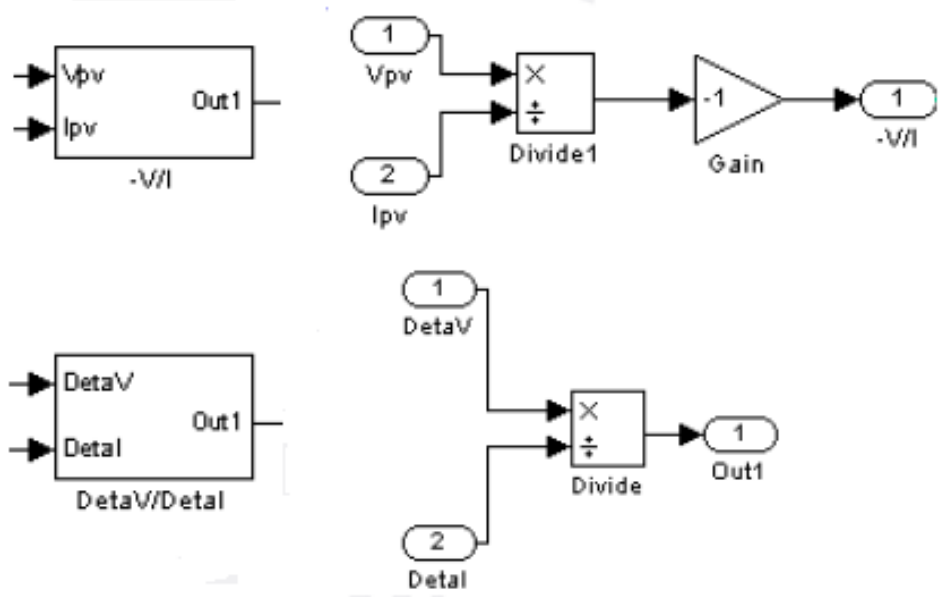

Fig. 9: Subblocks of -V/I and ΔV/ΔI

## III. BACKGROUND: POWER TRANSMISSION FROM SPACE

The sun shines continuously in space, so a PV array placed in orbit is not limited by the day-night cycle or by weather, which reduce the average daily period and intensity of insolation on the ground. The challenge is to transmit the collected energy to the earth. In a space-based solar power system, the collecting satellite converts solar energy into electrical energy and powers a microwave emitter aimed at a collector on the earth's surface.

Microwave beams have been proposed in preference to laser beams because of their higher efficiency. The primary components of a beamed microwave system are a microwave source, a transmitting antenna and a receiving rectenna, a rectifying antenna that converts the microwave energy directly into DC electricity. The rectenna was conceived and developed specifically for beamed microwave power transmission [15]. Because of diffraction, any beam spreads and weakens with distance, and the Rayleigh criterion implies that large transmitting and receiving apertures are required; this is one of the main disadvantages of the approach.

The detailed design of the transmission system is outside the scope of this paper, which concentrates on the PV array and MPPT simulation described in Sections II and IV.

## IV. SIMULATION RESULTS

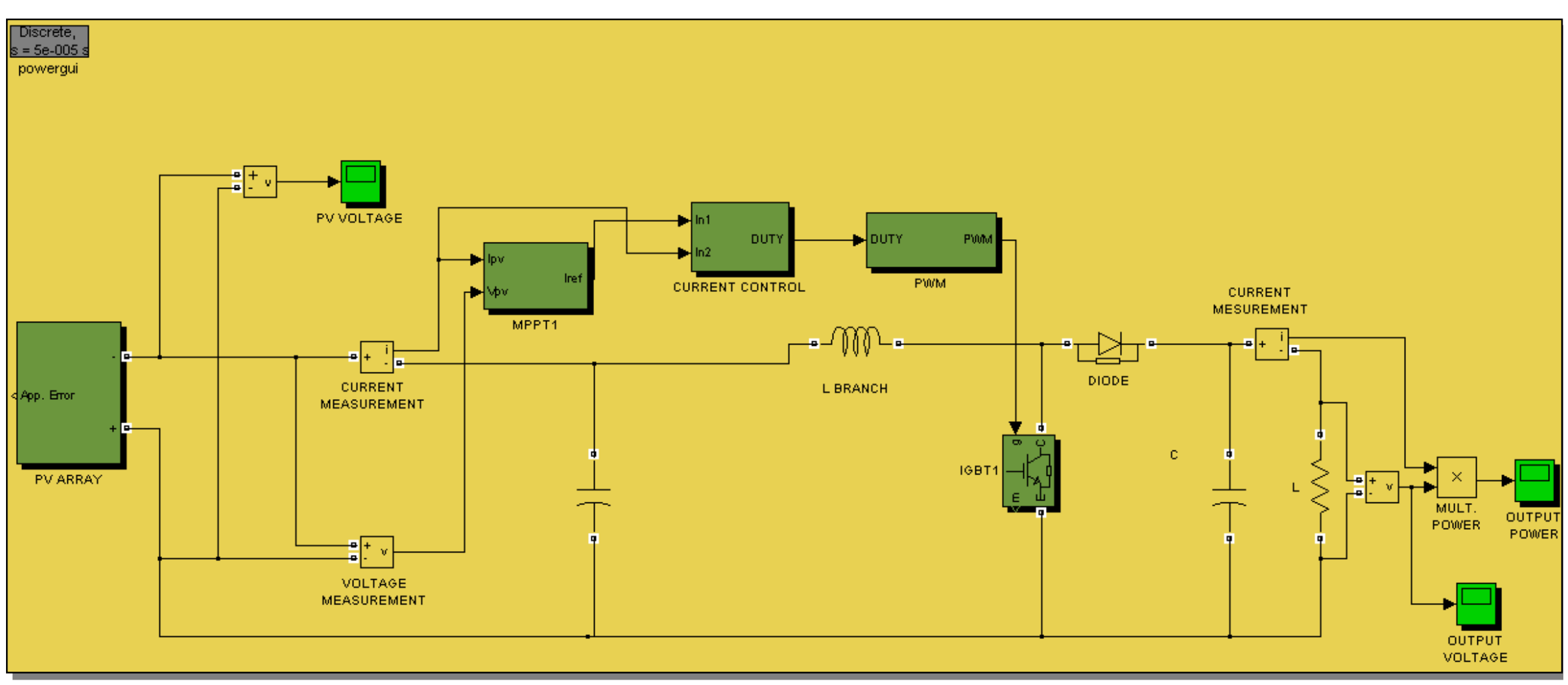

Fig. 10: MATLAB design with PV array, MPPT, power electronics, current controller and PWM

The building of the PV array and of the whole design is described in Section II. Here the required parameters, the simulation results and comments are presented. The following parameters were considered for the PV array.

| | |
|---|---|
| Number of Series Cells | 60 |
| Number of Parallel Cells | 1 |
| Series Resistor | 0.0005 |
| Thermal Voltage (Vt) | 0.02538 |
| Short Circuit Current (Isc) | 16.37 |
| Open Circuit Voltage (Voc) | 3.251 |
| Learning Coefficient | 0.0005 |
| Series Modules In Array | 2 |
| Parallel Modules In Array | 3 |
| | |
| *Efficiency of Cell* | *36.30%* |

Table 3: Block Parameters of PV Array

SIMULATION RESULTS *WITH MPPT*.

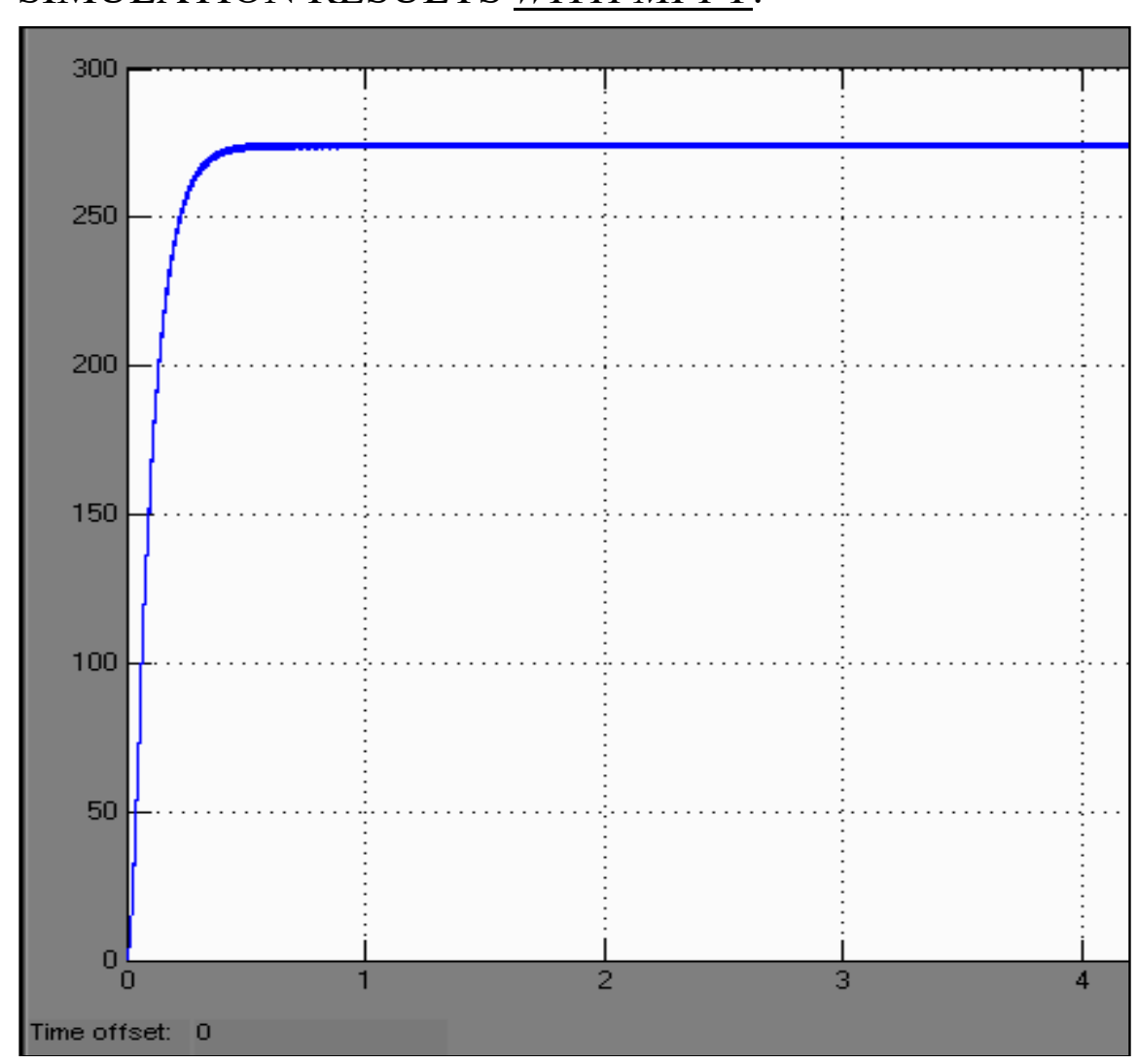

Fig. 11: Output Power of the System

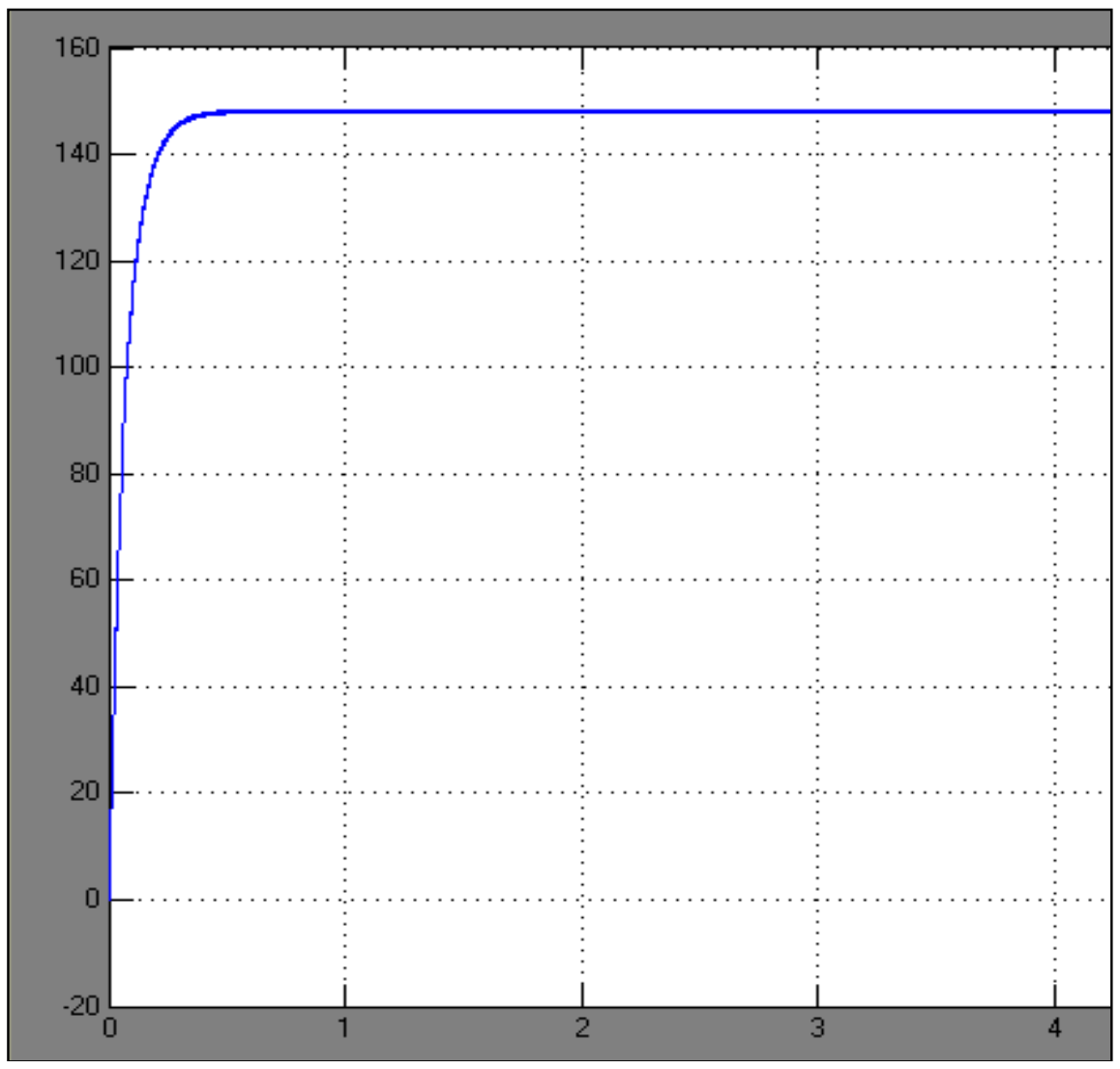

Fig. 12: Output Voltage of the System

SIMULATION RESULTS *WITHOUT MPPT.*

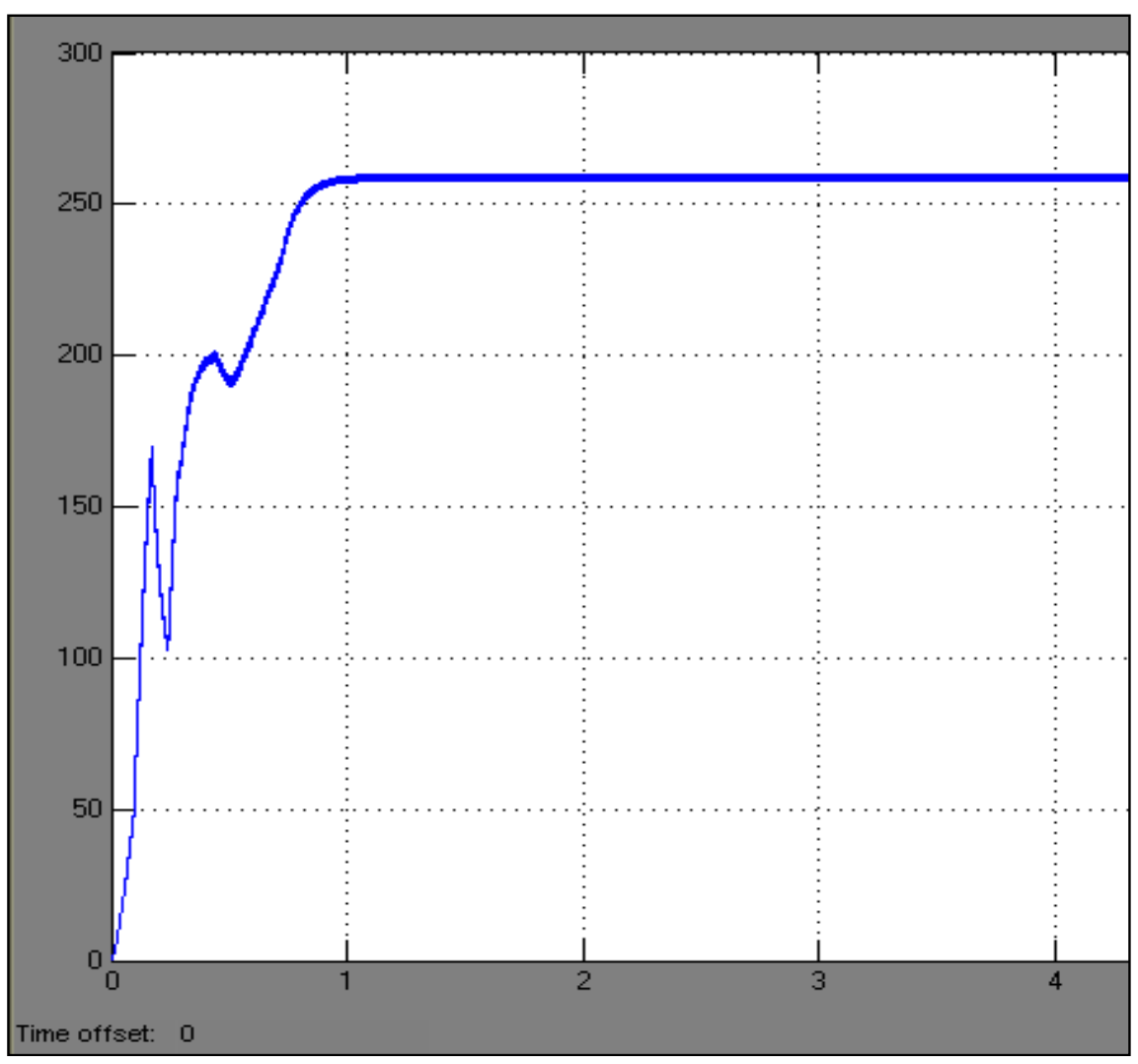

Fig. 13: Output Power of the System

With the consideration of the system as a whole, result with the MPPT application is better than the result of the system output power which is simulated without MPPT. But if we take a look to the output power, there is no big difference. If we compare it low efficient panel which is used ground solar application, for space applications high efficiency of the solar panels are decreasing the desired influence of the MPPT block.

Actually, in this simulation important part is; how the output responses at the very beginning of the simulation. System with MPPT has much better response to the demanded power at the beginning. Undesired fluctuations are almost eliminated with the MPPT.

IRRADIATION AT THE OUTER ATMOSPHERE OF THE EARTH AND ITS SIMULATION.

The solar irradiance outside the earth's atmosphere is relatively constant, about 1.36 kW/m². Unlike ground conditions, it is not affected by atmospheric absorption and scattering, clouds, water vapor, pollution, latitude, season or time of day. In the simulations the insolation is varied between 300 W/m² and 1353 W/m², where 1353 W/m² is used as the irradiance in space.

| *Insolation (W/m²)* | *I (A)* | *V (V)* | *Power (W)* | *Increment (coefficient)* |
|---|---|---|---|---|
| 300 | 0.50 | 26.50 | 9.00 | - |
| 500 | 0.85 | 27.50 | 17.00 | 1.7 |
| 700 | 1.15 | 28.00 | 25.00 | 1.352941 |
| 900 | 1.50 | 28.50 | 33.00 | 1.304348 |
| 1000 | 1.65 | 28.70 | 37.00 | 1.1 |
| 1100 | 1.80 | 28.90 | 40.10 | 1.090909 |
| 1300 | 2.10 | 29.00 | 49.00 | 1.166667 |
| 1353 | 2.20 | 29.10 | 51.00 | 1.047619 |

Table 4: Results with different insolation values

Comparing the array under ground-level and space insolation, the extracted power is higher at the higher insolation rate, as expected. The following observation is based on the simulation results (Table 4):

For the standard ground condition of 1000 W/m² the extracted power is 37 W; for the space insolation of 1353 W/m² it is 51 W, i.e. approximately 1.4 times higher.

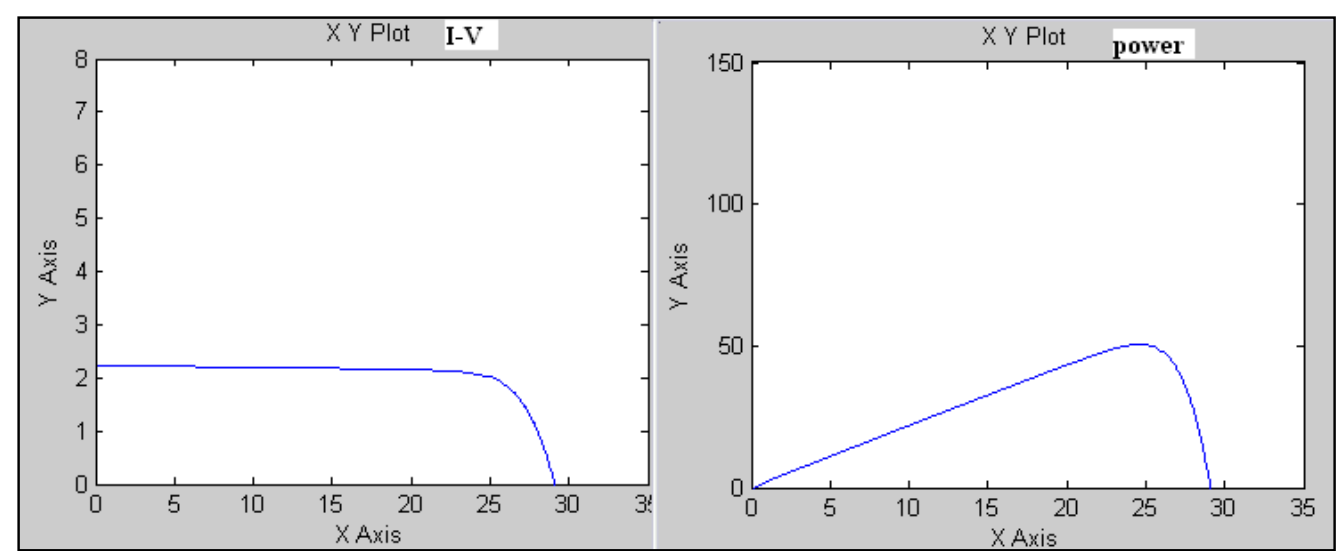

Fig. 14: Plots for insolation of 1353 W/m²

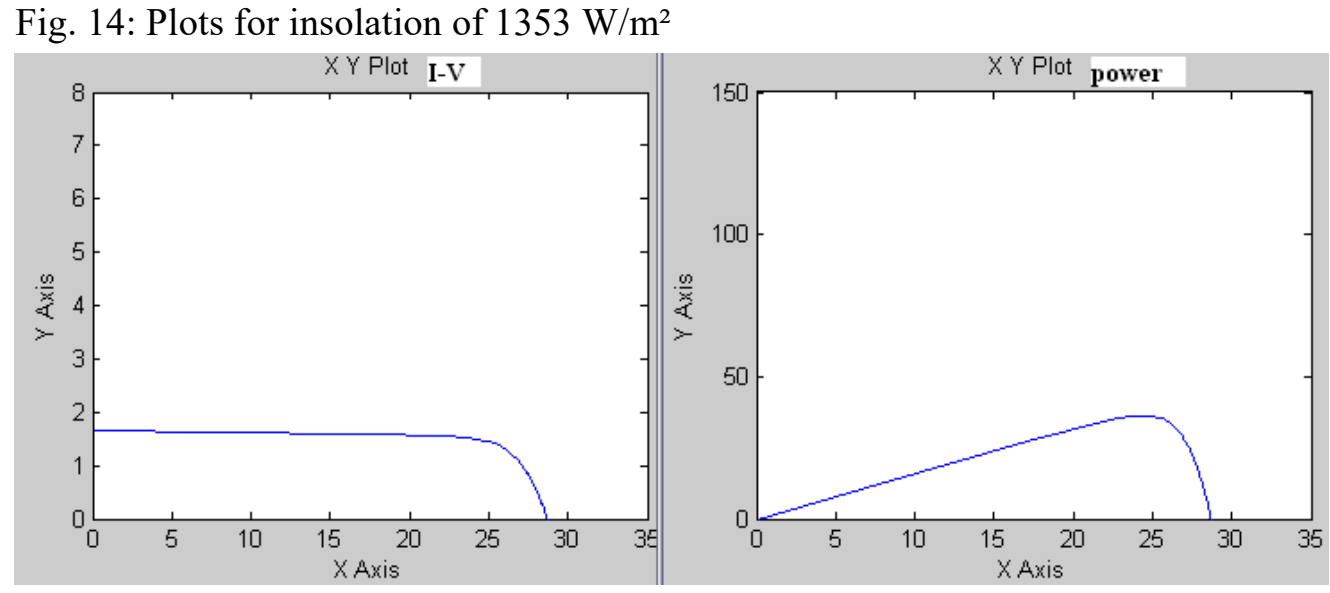

Fig. 15: Plots for insolation of 1000 W/m²

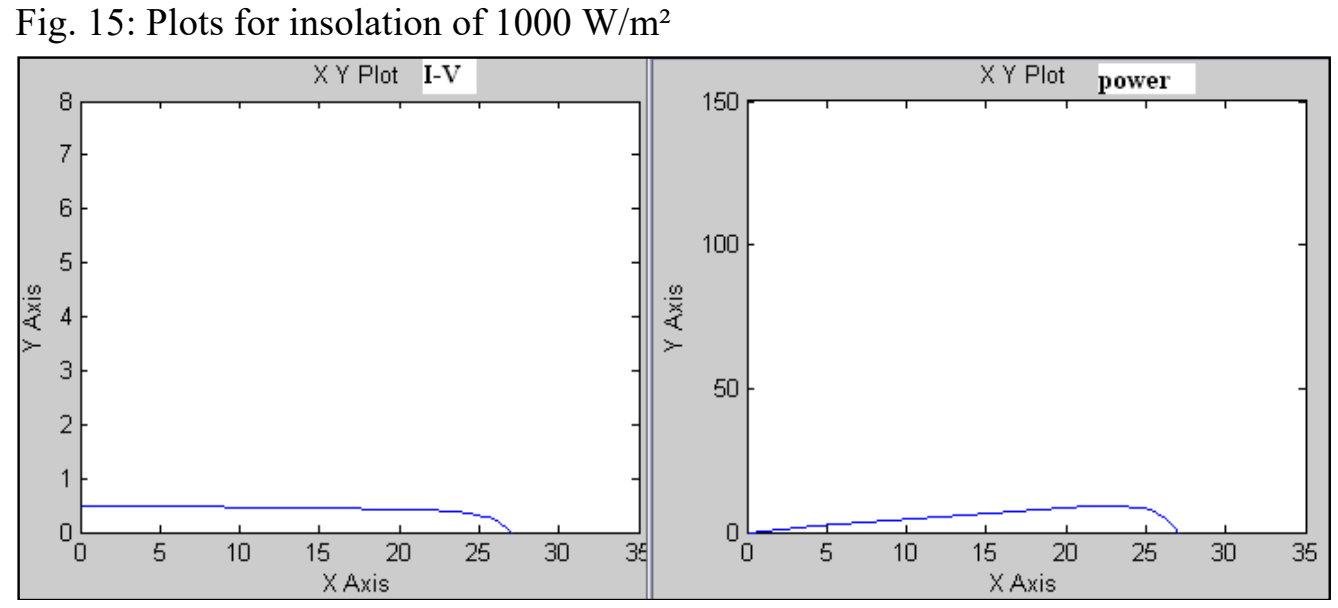

Fig. 16: Plots for insolation of 300 W/m²

The data collected for insolation between 300 W/m² and 1353 W/m² are given in Table 4.

## V. CONCLUSION

A PV array for space use was modeled in MATLAB/Simulink and simulated with and without incremental-conductance MPPT. The steady-state output power of the two cases differs little, because the high efficiency of the space-grade panel reduces the influence of the MPPT block. The benefit of MPPT appears in the response at the beginning of the simulation: the system with MPPT follows the demanded power much better, and undesired fluctuations are almost eliminated. Simulations at insolation levels between 300 W/m² and 1353 W/m² show that the extracted power increases with insolation; at the space value of 1353 W/m² it is approximately 1.4 times that at 1000 W/m² (51 W compared with 37 W).

The transmission of the generated power to the earth by a microwave beam was reviewed briefly; the large transmitting and receiving apertures it requires are its main drawback.

Challenges faced and problems solved:

During the group study, first of all we were not familiar with the MATLAB and Simulink environment, so the first challenge was to learn how to build the model from MATLAB online tutorials. After the model was set up, there were many modeling errors.

- The errors were mainly related to the Gain blocks: mismatched sample times caused the problem, which was solved by adding Unit Delay blocks at the stage following the Gain block.
- The PWM switching frequency was the other parameter, which was finally set to 10 kHz after many trials.
- The sawtooth signal and the duty signal could have been combined using logic and bit operations, but this was done with a relay block whose output is set to 1 when it is ON.
- Another issue was the sample time of the PI controller, which takes the difference between the reference current and the PV array current.
- Sample time for the PI Controller was set to 250E-6. The saturation limits for the calculation of the current change in the MPPT part are set to +15 and −15.